\documentclass[%
 aip,
 jmp,%
 amsmath,amssymb,
 preprint,%
author-numerical,
]{revtex4-1}
\newcommand{\be}{\begin{equation}}
\newcommand{\ee}{\end{equation}}

\newcommand{\ba}{\begin{array}}
\newcommand{\ea}{\end{array}}
\newcommand{\bea}{\begin{eqnarray}}
\newcommand{\eea}{\end{eqnarray}}
\newcommand{\nn}{\nonumber \\}

\usepackage{graphicx}
\usepackage{dcolumn}
\usepackage{bm}
\allowdisplaybreaks

\newcommand{\etaell}{{u}}
\newcommand{\xiell}{{v}}

\begin{document}


\title{Third-order superintegrable systems with potentials satisfying nonlinear equations }

\author{A. Marchesiello}
\email{anto.marchesiello@gmail.com}
\affiliation{Czech Technical University in Prague, \\
Faculty of Nuclear Sciences and Physical Engineering, \\
 B\v rehov\'a 7, 115 19 Prague 1, Czech Republic
}
\author{S. Post}%
 \email{spost@hawaii.edu}
\affiliation{
Department of Mathematics,
University of Hawai'i at M\=anoa\\
2625 McCarthy Mall, Honolulu (HI) 96822, USA
}%

\author{L. \v{S}nobl}\email{Libor.Snobl@fjfi.cvut.cz}

\affiliation{Czech Technical University in Prague, \\
Faculty of Nuclear Sciences and Physical Engineering, \\
 B\v rehov\'a 7, 115 19 Prague 1, Czech Republic
}
\date{\today}

\begin{abstract}
The conditions for superintegrable systems in two-dimensional Euclidean space admitting separation of variables in an orthogonal coordinate system and a functionally independent third-order integral are studied. It is shown that only systems that separate in subgroup type coordinates, Cartesian or polar, admit potentials that can be described in terms of nonlinear special functions. Systems separating in parabolic or elliptic coordinated are shown to have potentials with only non-movable singularities.
\end{abstract}

\maketitle
\title{\\
Draft \today}
\maketitle
 
\section{Introduction}
In their 2002 paper \cite{GW}, Gravel and Winternitz discovered new superintegrable potentials depending on solutions of nonlinear  ODEs including Painlev\'e transcendents and elliptic functions. These new systems are strikingly different from all previously known superintegrable systems. Indeed, while these systems are still exactly-solvable in the sense that their energy spectrum can be obtained  using algebraic methods, see e.g. \cite{MW2008, marquette2009painleve}, these Hamiltonians do not possess the stronger properties
usually associated with exact solvability. Namely, either that the Hamiltonian can be expressed as an element of the  enveloping algebra of a two-variable differential operator representation of a Lie algebra \cite{TempTW} or that the wavefunctions can be expressed as a ground-state multiplied by an orthogonal polynomial \cite{KMJP}. These new potentials also were superficially different in that they cannot be expressed as rational functions in any coordinate system.

The further investigation of  third-order superintegrable systems that admit separation of variables in Cartesian\cite{Gravel} and polar coordinates \cite{TW20101} lead to the discovery of even more ``nonlinear" potentials, with at least one potential depending on each of the Painlev\'e transcendents.  Here,  we emphasize that the potentials are solutions of nonlinear determining equations; the Schr\"odinger equation is still linear. These further discoveries, along with recalling that the defining equations for a superintegrable potential are nonlinear,
might lead one to believe that such ``nonlinear" potential are ubiquitous. However, extending
the search to other coordinate systems including parabolic\cite{PopperPostWint2012} and, as we shall see later, elliptic coordinates yields not only no nonlinear potentials but also no new potentials at all. Thus, we are left to conjecture that there is something special when subgroup-type coordinates are considered.

The goal of this paper is to review the previous results concerning  systems that admit a third-order integral and are separable in Cartesian, polar and parabolic coordinates as well as to present new results for the elliptic case. We shall show that nonlinear potentials are present only in the subgroup type coordinates while for the other two coordinate systems all potentials satisfy some set of linear ODEs. The mechanism for this disparity is still not well understood and so, in the conclusions, we offer some conjectures and open problems suggested by these results.

Before we specify the separation in a particular coordinate system, let us review the general determining equations for a third-order integral of motion, see e.g. \cite{GW}. We consider scalar Hamiltonians on 2D Euclidean space defined by
\be H=\frac{1}{2} (p_1^2+p_2^2)+V(x_1, x_2),\label{generalH}\ee

where either $p_j$ are the conjugate momentum for classical mechanics or $p_j=-i\hbar \partial_{x_j}$ for a quantum mechanical operator. Suppose now that the Hamiltonian admits an integral of the motion that is third order in the momenta. Due to the form of the Hamiltonian (\ref{generalH}), it is possible to express any third-order integral as
\be \label{Xintegral}
X=\frac12\left( \sum_{j,k,\ell}A_{j,k,\ell}\lbrace L_3^j,p_1^kp_2^\ell\rbrace + \lbrace g_1(x_1, x_2), p_1\rbrace +\lbrace g_2(x_1, x_2), p_2\rbrace\right),
\ee
where the bracket here is the symmetrizer $\{A, B\}=AB+BA$ which is trivial for $X$ a function on 4D phase space in classical mechanics
but necessary when the $p_j$ are differential operators. Also, $L_3=x_1p_2-x_2p_1$
is the generator of rotation. The integral (\ref{Xintegral}) contains only odd-order terms since even and odd order terms will commute independently.  See \cite{PostWintNth} for a proof of the generality of this form as well as the following determining conditions which ensure that $X$ is an integral of the motion. Namely, the determining conditions  reduce to the following PDEs. The highest order determining equations, obtained from fourth-order terms in the momenta are identically zero due to the choice of leading order terms in the integral  (\ref{Xintegral}), namely as elements of the enveloping algebra for $e_2(\mathbb{R})$ (and, vice versa, elements of the enveloping algebra are the only solutions of the highest order determining equation). The next set of determining equations are
obtained from requiring that terms that are second-order in the momenta are identically zero and are given by
\bea\label{g1det} &g_{1,1}=3F_1V_{,1}+F_2V_{,2}\\
 &g_{1,2}+g_{2,1}=2(F_2V_{,1}+F_3V_{,2})\\
&g_{2,2}=F_3V_{,1}+3F_4 V_{,2},\label{g2det}\eea
where
 \[ F_1\equiv -A_{300}x_2^3+A_{210}x_2^2-A_{102}x_2+A_{030},\]
\[ F_2\equiv 3A_{300}x_1x_2^2-2A_{210}x_1x_2+A_{201}x_2^2+A_{120}x_1-A_{111}x_2+A_{021},\]
\[ F_3\equiv -3A_{300}x_1^2x_2-2A_{201}x_1x_2+A_{210}x_1^2+A_{111}x_1-A_{102}x_2+A_{012},\]
\[ F_4\equiv A_{300}x_1^3+A_{201}x_1^2+A_{102}x_1+A_{003}.\]
The functions $F_j$ are the leading-order terms of $X$ as in
\[X= i \hbar^3  ( F_1\partial_1^3+F_2\partial_1^2\partial_2+F_3\partial_1\partial_2^2+F_{4}\partial_2^3)+\ldots.\]
In equations (\ref{g1det}-\ref{g2det}) and what follows, we use the notation
\[ f_{,i}=\partial_i f, \qquad \partial_i=\frac{\partial}{\partial x_i}.\]

The next set of determining { equations comes } from zeroth order terms in the commutator and are  given, modulo the other determining equations by
\bea \label{nonlindet} 0&=& \left(g_1-\hbar^2(-2A_{300}x_2+\frac12A_{210})\right)V_1
+\left(g_2-\hbar^2(2A_{300}x_1+\frac12 A_{201})\right)V_2\nn
&-& \frac{\hbar^2}{4}\left(F_1V_{111}+F_2V_{112}+F_3V_{122}+F_4V_{222}\right). \eea
The classical version is obtained by simply setting $\hbar=0$.  As discussed in \cite{PostWintNth}, the odd-ordered terms are differential consequences of these determining equations.

Given a potential $V(x_1, x_2)$, the functions $g_1$ and $g_2$ can be obtained immediately from the determining equations (\ref{g1det}-\ref{g2det}), which can then be substituted into (\ref{nonlindet}). The problem is then to find {\bf potentials } $V(x_1,x_2)$ and  leading-order {\bf constants  $\bf{A}$} that satisfy the compatibility conditions for these determining equations. The compatibility condition for (\ref{g1det}-\ref{g2det}) is
\bea \label{lincomp}
 0&=&-F_3V_{111}+(2F_2-3F_4)V_{112}+(-3{F}_1+2F_3)V_{122}-F_2V_{222}\nn
&+&2(F_{2,2}-F_{3,1})V_{11}+2(-3F_{1,2}+F_{2,1}+F_{3,2}-3F_{4,1})V_{12}+2(-F_{2,2}+F_{3,1})V_{22}\nn
&+&(-3F_{1,22}+2 F_{2,12}-F_{3,11})V_1+(-F_{2,22}+2F_{3,12}-3F_{4,11})V_2.\eea
(There are also nonlinear compatibility conditions that come from isolating  $g_1$ (assuming $V_1\ne 0$)  from the the fourth determining equation (\ref{nonlindet}), taking a derivative, using (\ref{g1det}-\ref{g2det}) to simplify, isolating $g_2$ and repeating the process. This results in three, nonlinear determining equations for the potential \cite{Gravel}.)

The remainder of the paper is dedicated to the solution of this problem, under the assumption that the potential separates in an orthogonal coordinate system in real 2D Euclidean space. Under this assumption, the potential $V(x_1, x_2)$ can be written in terms of two { one--variable} functions. The linear compatibility conditions (\ref{lincomp}) can then be reduced to a pair of {  ODEs}. In particular, by assuming that $V(x_1,x_2)$ can be expressed in terms of the new potentials $V_1(y_1)$ and $V_2(y_2)$ and restricting one of the variables to a regular point, for example $y_2=y_2^{0}$, the linear compatibility condition is reduced to a linear, third-order ODE for $V_1(y_1)$. Thus, unless (\ref{nonlindet}) is satisfied identically, the potential will have only non-movable singularities. We call such potentials ``linear." In this case, the potentials satisfy also the nonlinear compatibility conditions but the equations act instead by reducing the solution space of the linear equations.  On the other hand, if (\ref{lincomp}) is identically zero, then the potential will be a solution to the nonlinear compatibility conditions, and may have movable singular points. It is these solutions that correspond to nonlinear special functions, including Painlev\'e transcendents and elliptic functions.

 Section II gives the analysis for subgroup-type coordinates (Cartesian and polar) as well as the resulting ``nonlinear" potentials. Section III gives the analysis for the remaining coordinate systems (parabolic and elliptic), including the result that all potentials are ``linear" and in particular have only non-movable singular points. Section IV includes some concluding remarks and conjectures.

\section{Subgroup-type Coordinates}
As it is well known, there are 4 in-equivalent choices of orthogonal coordinates for which the Helmholtz (zero-potential) admits separation of variables\cite{eisenhart1948enumeration, Miller}. Namely, Cartesian, polar, parabolic and elliptic coordinates.
Before presenting the results on the potentials, let us briefly describe the group theoretic interpretation of the separable coordinates in 2D real Euclidean space. Each coordinate system is determined by a second-order linear operator that commutes with the Laplacian. Up to conjugation by elements of the group $e_2(\mathbb{R})$, the second order operators are
\[ p_1^2, \qquad L_3^2, \qquad p_1L_3, \qquad L_3^2+c p_1^2.\]
The first two, $p_1^2$ and $L_3^2$ are second-order invariants of the two, inequivalent maximal Lie subalgebras of $e_2(\mathbb{R})$, namely the subalgebra generated by $\{p_1, p_2\}$ and $\{L_3\}$ respectively. Thus, these coordinates, Cartesian and polar respectively, are referred to  as subgroup type coordinates\cite{miller1981subgroups}.

\subsection{Cartesian}
We begin with Cartesian coordinates and summarize the results of Gravel\cite{Gravel}. Setting $x_1=x$, $x_2=y$, the potential separates as $V(x,y)=V_1(x)+V_2(y),$
so that the Hamiltonian is given by
\[H=\frac{-\hbar^2}{2}\left[\partial_x^2+\partial_y^2\right]+V_1(x)+V_2(y),\]
with second-order integral
\[Y=\frac{-\hbar^2}{2}\partial_x^2+V_1(x).\]
 The linear compatibility condition \eqref{lincomp} { reduces } to
\bea -F_{3}V_{1,xxx}-4F_{3,x}V_{1,xx}-6F_{3,xx}V_{1,x}= F_2V_{2,yyy}+ 4 F_{2,y}V_{2,yy}+6 F_{2,yy}V_{2y}.\label{cccart}\eea
Here the determining equations for the coefficient functions $F_j$, namely
$$F_{j,x}+F_{j-1,y}=0, \qquad j=1\ldots 5, \qquad F_{0}=F_{5}=0,$$
have been used to simplify (\ref{cccart}). { Fixing the value of $y$ at a regular point $y^0$}, the compatibility condition (\ref{cccart}) becomes a third-order ODE for the function $V_1$ with only regular singular points, since the leading order term is a  quadratic polynomial in $x$.  Similar reasoning shows that $V_2$ has only regular singular points.  Thus, if both sides of equation (\ref{cccart}) are not identically zero the potential has only regular singular points which are of course non-movable.

Let's consider the potentials (and coefficients $A_{jk\ell}$) for which at least one of the sides of equation (\ref{cccart}) vanishes identically. There are  two possible cases. The first is that (\ref{cccart}) is identically satisfied for any $V_1$ and $V_2$ by choosing appropriate coefficients. This would be the case if and only if both $F_3$ and $F_2$ were identically 0. The second case is where $F_3$ is identically 0 and $V_2$ satisfies a linear ODE, or equivalently $F_2$ is identically 0 and $V_1$ satisfies a linear ODE.
\paragraph{Case 1} Let us assume the coefficients are chosen so that $F_3\equiv 0$ and $F_2\equiv 0$. In this case, the only non-zero $A_{jk\ell}$ are $A_{030}$ and $A_{003}$ and it is possible to solve directly for $g_1(x,y)$ and $g_2(x,y)$  and to substitute these solutions in (\ref{nonlindet}). By separating variables and integrating, the nonlinear determining equations for $V_1(x)$ and $V_2(y)$ are reduced to
\bea \hbar^2V''_1(x)=6V_1(x)^2+A_{003}\sigma x, \\
         -\hbar^2V''_2(y)=6V_2(y)^2+A_{030}\sigma y.\eea
        Depending on whether or not $\sigma=0$, the solutions are either Weierstrass elliptic functions or Painlev\'e transcendents $P_1$. In the classical case, $\hbar=0$ and the potentials are linear combinations of $\pm \sqrt{x}$ and $\pm \sqrt{y}.$

\paragraph{Case 2} The other possibility is for the linear compatibility equations for $V_1(x)$ to be satisfied identically and the one for $V_2(y)$ to be non-trivial. This is the case when $F_3\equiv 0$ so that the only non-zero coefficients are $A_{030}, A_{003},  A_{120}$ and $A_{021}$.
In this case $F_2=A_{120}x+A_{021}$ { } and  $V_2(y)$ satisfies the linear ODE
\[0=(A_{120}x+A_{021})V_2'''\Rightarrow V_2(y)=ay^2+a'y.\]
Thus, up to translation in $y$, there are only two distinct cases $V_2(y)=ay^2$ and $V_2(y)=ay.$ Recall, a constant term in the potential is trivial. In the former case, new quantum solutions { for $V_1(x)$} are obtained that depend on the fourth Painlev\'e transcendent and new classical solutions are obtained
as a root of a fourth-order polynomial equation, and includes a smooth interpolation between a 1:1 and 3:1 anisotropic oscillator. For the final family of solutions, with ${ V_2(y)}=ay$, new solutions are obtained depending on, again, the first Painlev\'e transcendent as well as new solutions depending on the second Painlev\'e transcendent.

To briefly review the Cartesian case, not only were new linear potentials obtained, such as the 3:1 oscillator and its singular form,  but also new nonlinear potentials in both classical and quantum mechanics. New quantum potentials included ones depending on the first, second and fourth Painlev\'e transcendents.

\subsection{Polar}
Let us now consider systems that admit separation of variables in the polar coordinates
\[x_1=r\cos \theta, \qquad x_2=r\sin \theta,\]
and so the Hamiltonian  can be expressed as
\[H=\frac{-\hbar^2}{2}\left(\frac1r \partial_r r\partial_r +\frac{1}{r^2}\partial_\theta^2\right)+R(r)+\frac{1}{r^2}S(\theta).\]
The second-order integral of motion responsible for separation of variables is
\[Y=\frac{-\hbar^2}{2}\partial_\theta^2+S(\theta).\]
The analysis of this case was first performed by Tremblay and Winternitz\cite{TW20101} and here we review the results of their analysis.
In polar coordinates, the functions $F_j$ serve the same function as in the Cartesian case, namely that they are the leading terms of the 3rd-order integral expanded out with coefficient functions written on the left,
\[X=i\hbar^3\left(F_4\partial _\theta^3+F_3\partial_\theta^2\partial_r+F_2\partial_\theta\partial_r^2+F_1\partial_r^3\right)+\ldots, \]
The functions $F_j$ are defined as
\bea \ba{rl} F_1&=A_1\cos 3\theta +A_2\sin 3\theta +A_3 \cos \theta +A_4\sin\theta\\
F_2&=\frac{-3A_1\sin 3\theta+3A_2\cos 3\theta -A_3\sin \theta +A_4\cos \theta}{r}+B_1\cos 2\theta +B_2\sin 2 \theta+B_0\\
F_3&= \frac{-3A_1\cos 3\theta-3A_2\sin 3\theta +A_3\cos \theta +A_4\sin \theta}{r^2}+\frac{-2B_1\sin 2\theta +2B_2\cos 2 \theta}{r}+C_1\cos\theta +C_2\sin \theta\\
F_4&=\frac{A_1\sin 3 \theta -A_2\cos 3\theta -A_3\sin \theta +A_4 \cos \theta}{r^3} -\frac{B_1\cos 2\theta+B_2\sin 2 \theta  - B_0}{r^2}-\frac{C_1\sin\theta-C_2\cos\theta}{r}+D_0. \ea \nonumber\eea
The coefficients are related to those in the previous section via
\bea \ba{llll}A_1=\frac{A_{030}-A_{012}}{4}, &A_2=\frac{A_{021}-A_{003}}{4},& A_3=\frac{3A_{030}+A_{012}}{4},&\\
A_4=\frac{3A_{003}+A_{021}}{4}, & B_1=\frac{A_{120}-A_{102}}{2},&B_2=\frac{A_{111}}2, & B_0=\frac{A_{120}+A_{102}}2,\nn
C_1=A_{210}, & C_2=A_{201}, & D_0=A_{300}& \ea\nonumber\eea
as in the original paper of Tremblay and Winternitz\cite{TW20101}. The equations (\ref{g1det}-\ref{g2det}) are equivalently transformed. Most important for our analysis is the form of the linear compatibility conditions, which can be expressed as
\bea  \label{ccpolar} &r^4F_3R'''+r\left(4r^3F_{3,r}+6r^2F_3+3F_1\right)R''+\left(6r^4F_{3,rr}+20r^3F_{3,r}+6r^2F_{3}- 3 F_1\right)R'=\\
&\qquad-r^{-2}\left(F_2S'''+4F_{2,\theta}S''+(6F_{2,\theta \theta}-6F_{2,r}r+4F_{2})S'+\left(12rF_{2,\theta r}-8F_{2,\theta}\right)S\right)+36r^{-3}F_{1}S.\nonumber\eea
Again, we have used the first-order PDEs for the coefficient functions $F_j$ to simplify this expression.

In this form, it is clear to see that a regular point $\theta=\theta^{0}$ (\ref{ccpolar}) gives a linear ODE for $R(r)$ and so the solutions will have only fixed singular points as long as the left-hand side of (\ref{ccpolar}) is not identically satisfied. This ODE will be identically satisfied if and only if $F_3\equiv 0$ and $F_1\equiv 0$ or equivalently $A_1=A_2=A_3=A_4=0$ , $B_1=B_2=0$ and $C_1=C_2=0$. In this case the only non-zero coefficients are $B_0=(A_{120}+A_{102})/2$ and $D_0=A_{300}$. 
With these values of the coefficients, the problem splits into two cases. When $B_0\neq 0$ the potential is purely radial and the remaining conditions force the third order integral $X$ to be algebraically dependent on the integrals $L_3$ and $H$. Thus the potential is not superintegrable but first order integrable. When $B_0=0$ , the right-hand side of (\ref{ccpolar}) is also identically 0 and so the linear compatibility conditions for $R(r)$ and $S(\theta)$ are both identically satisfied. This is the analog of Case 1 in the Cartesian case.

For the other case, let's assume that $R(r)$ satisfies a non-trivial linear ODE and so has only non-movable singularities. Again, $S(\theta)$ will also have only non-movable singularities unless the right-hand side of (\ref{ccpolar}) is identically satisfied. This will be the case as long as $A_1=A_2=A_3=A_4=0$  and $B_1=B_2=B_0=0. $ Unlike the previous case, the values of $C_1$ and $C_2$ maybe non-zero. This is Case 2.

 Note that unlike separation in Cartesian coordinates, there is asymmetry between the functions $R(r)$ and $S(\theta). $ If the linear compatibility conditions for $R(r)$ are satisfied identically then those for $S(\theta)$ will be as well. This is due to the fact that the functions $F_3$ and $F_1$ depend on more of the constants $A_{jk\ell}$ than $F_2$ and so $F_3\equiv 0$ (together with the assumption that $X$ is an independent third integral) implies $F_2\equiv 0$ but the other direction does not hold. As we shall see, there will be solutions where $R(r)$ satisfies a linear ODE while $S(\theta)$  satisfies a nonlinear one.

\paragraph{Case 1 Polar} In case 1 for polar potentials all of the $A_{jk\ell}$ are 0 except  $A_{300}$ and $A_{120}+A_{102}$. From direct computations, the choice $A_{120}+A_{102}\ne 0$ leads to known potentials and the only new case is $A_{300}\ne 0.$  In this case, the angular part of the potential can be expressed as a Weierstrass elliptic function. However, the third-order integral obtained will be algebraically dependent on the second-order integral responsible for separation of variables. Thus this system does not admit an algebraically independent third-order integral and is not considered in this classification.
\paragraph{Case 2 Polar} Here we assume that the linear conditions for the angular part of the potential are satisfied identically.  The only non-zero constants are $C_1, C_2$ and $D_0$ and the radial term of the potential is given, up to a trivial additive constant,  by
\[R(r)=\frac{a_1}{r}+\frac{a_2}{r^2}.\]
In the original paper of Tremblay and Winternitz\cite{TW20101}, it is shown that the only potential that survives the remaining determining equations (\ref{g1det}-\ref{g2det}) and (\ref{nonlindet})  with $a_1\ne 0$ is the singular Coulomb potential, a well-know second-order superintegrable systems. When $a_1=0$ the term $a_2/r^2$ may be incorporated into the angular part of the potential and so we are left with the case that $R(r)=0$. In this case, there is a single nonlinear potential in classical mechanics, which satisfies a first-order nonlinear ODE, and two potentials in quantum mechanics one depending on the sixth Painlev\'e transcendent and the second depending on the Weierstrass elliptic function.

To briefly review the results of the polar coordinate case, there exist three ``nonlinear" potentials, one in classical and two in quantum mechanics. Each of these three nonlinear potentials arise only in the angular part.

\section{Non-subgroup-type Coordinates}
\subsection{Parabolic}
Turning now to the conditions that the potential separates in parabolic coordinates
\be \label{paraboliccords} x_1=\frac{\xi^2-\eta^2}{2}, \qquad x_2=\xi\eta,\ee
and with the Hamiltonian  of the form
\be \label{parabolicV} H=\frac{-\hbar^2}{ 2(\xi^2+\eta^2)}\left[\partial_\xi^2+\partial_\eta^2\right]+\frac{W_1(\xi)+W_2(\eta)}{\xi^2+\eta^2}.\ee
The second-order integral responsible for separation of variables is given by
\[ Y=\frac12 \{ p_2, L_3\} +\frac{\xi^2 W_2(\eta)-\eta^2 W_1(\xi)}{\xi^2+\eta^2},\]
with
\[p_2=\frac{-i\hbar}{\xi^2+\eta^2}\left[\xi\partial_\eta + \eta\partial_\xi\right], \qquad L_3=\frac{i\hbar}{2}\left[\xi \partial_\eta-\eta\partial_\xi\right].\]
The conditions that the potential \eqref{generalH} with a potential of this form admits a third-order integral were investigated by an author along with Popper and Winternitz\cite{PopperPostWint2012}. Unlike the Cartesian and polar cases,  it was shown that there are no potentials that satisfy nonlinear determining equations. In this section we review these results and recast the determining equations in an analogous way with the previous sections.

As in previous cases, we define functions $F_j$  composing the leading terms of the 3rd-order integral expanded out with coefficient functions written on the left,
\[X=i\hbar^3\left(F_1\partial _\xi^3+F_2\partial_\xi^2\partial_\eta+F_3\partial_\xi\partial_\eta^2+F_4\partial_\eta^3\right)+ \ldots .\]
\bea F_1&=&-\frac{\eta^3A_{300}}{8}+\frac{\eta^2(\xi A_{210}+\eta A_{201})}{4(\xi^2+\eta^2)}-\frac{\xi^2\eta A_{120}+\eta^2\xi A_{111}+\eta^3A_{102}}{2(\xi^2+\eta^2)^2}\nonumber\\
&&+\frac{\xi^3A_{030}+\xi^2\eta A_{021}+\eta^2\xi A_{012}+\eta^3A_{003}}{(\xi^2+\eta^2)^3}\nonumber\\
F_2&=&\frac{3\eta^2\xi A_{300}}{8}-\frac{\eta(\eta^2+2\xi^2)A_{210}+\eta^2 \xi A_{201}}{4(\xi^2+\eta^2)}+\frac{\xi(2\eta^2+\xi^2)A_{120}+\eta^3 A_{111}-\eta^2\xi A_{102}}{2(\xi^2+\eta^2)^2}\nonumber\\
&&-\frac{3\xi^2\eta A_{030}+\xi(2\eta^2-\xi^2) A_{021}+\eta(\eta^2-2\xi^2)A_{012}-3\eta^2\xi A_{003}}{(\xi^2+\eta^2)^3}\nonumber\\
F_3&=&-\frac{3\eta\xi^2 A_{300}}{8}+\frac{\xi (2\eta^2+\xi^2) A_{210}-\eta \xi^2 A_{201}}{4(\xi^2+\eta^2)}
+\frac{\xi^3A_{111}+\eta\xi^2A_{102}-\eta(\eta^2+2\xi^2)A_{120}}{2(\xi^2+\eta^2)^2}\nonumber\\
&&+\frac{3\eta^2\xi A_{030}+\eta(\eta^2-2\xi^2) A_{021}+\xi(\xi^2-2\eta^2)A_{012}+3\eta\xi^2 A_{003}}{(\xi^2+\eta^2)^3}\nonumber\\
 F_4&=&
\frac{\xi^3A_{300}}{8}+\frac{\xi^3 A_{201}-\eta\xi^2 A_{210}}{4(\xi^2+\eta^2)}+\frac{\eta^2\xi A_{120}-\eta\xi^2 A_{111}+\xi^3A_{102}}{2(\xi^2+\eta^2)^2}\nonumber\\
&&+\frac{\xi^3A_{003}+\eta^2\xi A_{021}-\eta\xi^2 A_{012}-\eta^3A_{030}}{(\xi^2+\eta^2)^3}.\nonumber
\eea
Note that these are not the same $F_j$ given in a previous paper\cite{PopperPostWint2012}, we have chosen this for consistency with the other cases of this article.  Assuming the potential separates as \eqref{parabolicV}, the linear determining equation becomes
\bea &&F_3W_{1}'''+\left(4F_{3,\xi}+(F_3+3F_1)\xi(\xi^2+\eta^2)^{-1}\right)W_1'' \nonumber\\&&
+\left(6F_{3,\xi\xi}+(\xi^2+\eta^2)^{-1}\left(6\xi F_{3,\xi}-6\eta  F_{3,\eta}+12\xi F_{1,\xi}-3(F_3-3F_1)\right)\right)W_1' +C_1 W_1=\nonumber\\
&&-F_2W_{2}'''-\left(4F_{2,\eta}+(F_2+3F_4)\eta(\xi^2+\eta^2)^{-1}\right)W_2'' \nonumber\\&&
-\left(6F_{2,\eta \eta}+(\xi^2+\eta^2)^{-1}\left(6\left(\eta F_{2,\eta}-\xi F_{2,\xi}+2\eta F_{4,\eta}\right)-3(F_2-3F_4)\right)\right)W_2'-C_2W_2, \label{paraboliclinearcomp}
\eea
with
\bea
C_1&=&-\frac{12\xi F_{3,\xi\xi}-12\eta F_{3,\eta \xi}+12\eta^{-1}(2\eta^2-\xi^2)F_{1,\xi \eta}}{\xi^2+\eta^2} \nonumber\\&
&-\frac{24(\xi^2-\eta^2)F_{3,\xi}-24\xi\eta F_{3,\eta}+12\eta^{-2}(4\eta^4-2\eta^2\xi^2+\xi^4)F_{1,\xi}+12(4\eta^2-3\xi^2)\xi\eta^{-1}F_{1,\eta}}{(\xi^2+\eta^2)^2}\nonumber \\&&
-\frac{12\xi \left(2\eta^4F_3+3(\xi^4-\xi^2\eta^2)F_1\right)}{\eta^{2}(\xi^2+\eta^2)^{3}}, \nonumber \eea
and
\bea C_2&=&-\frac{12\eta F_{2,\eta\eta}-12\xi F_{2,\eta \xi}+12\xi^{-1}(2\xi^2-\eta^2)F_{4,\xi \eta}}{\xi^2+\eta^2} \nonumber\\&
&+\frac{24(\xi^2-\eta^2)F_{2,\eta}+24\xi\eta F_{2,\xi}-12\xi^{-2}(4\xi^4-2\eta^2\xi^2+\eta^4)F_{4,\eta}-12(4\xi^2-3\eta^2)\xi^{-1}\eta F_{4,\xi}}{(\xi^2+\eta^2)^2}\nonumber \\&&
-\frac{12\eta \left(2\xi^4F_2+3(\eta^4-\xi^2\eta^2)F_4\right)}{\xi^{2}(\xi^2+\eta^2)^{3}}.\nonumber \eea

From the linear compatibility equation, it is clear that both of the functions in the potential will have only non-movable singularities unless either the functions $F_3$ or $F_2$ are identically 0. In either case, this would require all of the $A_{j k \ell}$ to vanish and so there would not be a third-order integral of motion. Thus, all potentials that separate in parabolic coordinates and admit a third-order integral of motion, depend on two functions that satisfy third-order linear ODEs.

 In practice, it is preferable to not deal directly with \eqref{paraboliclinearcomp} at regular points to obtain the separated equations for each of the components of the potential. Instead,  repeated derivatives of  \eqref{paraboliclinearcomp} were computed until the equation depended only on one function, for example $W_1(\xi)$ and then the coefficients of powers of $\eta$ gave a set of linear ODEs for $W_1(\xi)$. These ODEs were solved \cite{PopperPostWint2012} and, somewhat surprisingly, all the potential obtained were already known and admitted a third, second-order integral of motion beyond the Hamiltonian and the integral associated with separation of variables in the coordinates \eqref{paraboliccords}.

To review, unlike the subgroup-type coordinates, the linear compatibility conditions for a third-order integral for a potential that separates in parabolic coordinates is never identically satisfied and so all such potentials have only non-movable singularities. Furthermore, all such potentials are actually second-order superintegrable and the third-order integral is obtained as the commutator of the two, non-commuting second-order ones.

\subsection{Elliptic}

Let us consider elliptic coordinates in the form
\be\label{elliptic coordinates}
\left\{
  \begin{array}{ll}
    x_1&=\etaell \xiell \\ [2mm]
    x_2 &= \sqrt{1-\etaell^2}\sqrt{\xiell^2-1}
  \end{array}
\right.
\ee
where the ranges for the coordinates are $-1\leq\etaell\leq1$, $\xiell\geq1$. They cover the Cartesian halfplane prescribed by the condition $x_2\geq 0$.

The Hamiltonian separates in these coordinates if it is given by
 \be\label{Hamiltonian-elliptic}
 H=-\frac{\hbar ^2(\etaell\partial_{\etaell}+(\etaell ^2-1)\partial_{\etaell}^2-\xiell  \partial_{\xiell}-(\xiell ^2-1) \partial_{\xiell}^2)}{2 \left(\etaell ^2-\xiell ^2\right)}+ \frac{W_1(\etaell)+W_2(\xiell)}{\etaell^2-\xiell^2}.
 \ee
This form of the Hamiltonian implies the existence of a second order integral responsible for the separation in the elliptic coordinates~\eqref{elliptic coordinates}
\be
Y=L_3^2+\frac12(p_1^2-p_2^2)+ F(\etaell, \xiell),
\ee
with
\bea
p_1&=&i\hbar\frac{\xiell\left(1-\etaell ^2\right)\frac{\partial}{\partial \etaell}+\etaell  \left(\xiell ^2-1\right) \frac{\partial}{\partial \xiell }}{\etaell ^2-\xiell ^2} \nn
p_2&=&-i\hbar\frac{\sqrt{(1-\etaell ^2) (\xiell^2-1)}}{\etaell^2-\xiell^2}\left(\etaell\frac{\partial}{\partial \etaell}-\xiell\frac{\partial}{\partial \xiell }\right) \nn
L_3&=&-i\hbar\frac{\sqrt{(1-\etaell ^2)(\xiell^2-1)}}{\etaell ^2-\xiell ^2}\left(\xiell \frac{\partial}{\partial \etaell }-\etaell  \frac{\partial}{\partial \xiell }\right) \nonumber
\eea
and
$$ F(\etaell, \xiell) = \frac{(2\xiell^2-1)W_1(\etaell)+(2\etaell^2-1)W_2(\xiell)}{\etaell^2-\xiell^2}.$$

Written in the elliptic coordinates \eqref{elliptic coordinates}, the linear compatibility condition reads
\bea \label{linearcompell} &&
\frac{F_3(1-\etaell^2)}{\xiell^2-1} W_1'''+\left[\frac{4(1-\etaell^2)F_{3,\etaell}}{\xiell^2-1}-\frac{\etaell(F_3-3F_1)}{\etaell^2-\xiell^2}\right]W_1''\nonumber\\&&
+\left[\frac{6F_{3,\etaell \etaell}(1-\etaell^2)}{\xiell^2-1}+\frac{-2\etaell(2\etaell^2+\xiell^2-3)F_{3,\etaell}+6(\xiell^2-1)(\xiell F_{3,\xiell}+2\etaell F_{1, \etaell})}{(\xiell^2-1)(\etaell^2-\xiell^2)}\right.\nonumber\\&& \qquad
\left.+\frac{(\etaell^2\xiell^2+11\etaell^2-9\xiell^2-3)F_3+3(5\etaell^2+3)(\xiell^2-1)F_1}{(1-\etaell^2)(\xiell^2-1)(\etaell^2-\xiell^2)}\right]W_1'+C_1W_1= \nonumber\\
&&
-\frac{F_2(\xiell^2-1)}{1-\etaell^2} W_2'''-\left[\frac{4(\xiell^2-1) F_{2,\xiell}}{1-\etaell^2}+\frac{\xiell(F_2-3F_4)}{\etaell^2-\xiell^2}\right]W_2''\nonumber\\&&
-\left[\frac{6F_{2,\xiell \xiell}(\xiell^2-1)}{1-\etaell^2}-\frac{2\xiell(\etaell^2+2\xiell^2-3)F_{2,\xiell}+6(1-\etaell^2)(\etaell F_{2,\etaell}+2\xiell F_{4, \xiell})}{(1-\etaell^2)(\etaell^2-\xiell^2)}\right.\nonumber\\&& \qquad
\left.-\frac{(\etaell^2\xiell^2+11\xiell^2-9\etaell^2-3)F_2-3(5\xiell^2+3)(1-\etaell^2)F_4}{(1-\etaell^2)(\xiell^2-1)(\etaell^2-\xiell^2)}\right]W_2'-C_2W_2.
\eea
with
\bea {C}_1 &=&\frac{-12\etaell(1-\etaell^2) F_{3,\etaell \etaell}}{(\xiell^2-1)(\etaell^2-\xiell^2)}-\frac{12\xiell F_{3, \xiell \etaell}}{\etaell^2-\xiell^2}
+\frac{4(4\etaell^4+7\etaell^2\xiell^2+\xiell^4-6\etaell^2-6\xiell^2)F_{3,\etaell}}{(\etaell^2-\xiell^2)^2(\xiell^2-1)}\nonumber\\ &&
+\frac{12\xiell\etaell(\etaell^2+\xiell^2-2)F_{3, \xiell}}{(\etaell^2-\xiell^2)^2(1-\etaell^2)} +
\frac{4\etaell\left(\etaell^4\xiell^2-8\etaell^2\xiell^4+\xiell^6+5\etaell^4-4\etaell^2\xiell^2+11\xiell^4-6\xiell^2 \right)F_3}{(\etaell^2-\xiell^2)^3(1-\etaell^2)(\xiell^2-1)}\nonumber\\ &&
+\frac{12(2\etaell^2\xiell^2+\xiell^4-\etaell^2-2\xiell^2)F_{1,\xiell\etaell}}{ (\etaell^2-\xiell^2)(1-\etaell^2)\xiell}\nonumber\\ &&
-\frac{12(2\etaell^4\xiell^4+4\etaell^2\xiell^6+\xiell^8-2\etaell^4\xiell^2-8\etaell^2\xiell^4-4\xiell^6+\etaell^4+2\etaell^2\xiell^2+4\xiell^4)F_{1, \etaell}}{(\etaell^2-\xiell^2)^2(1-\etaell^2)(\xiell^2-1)\xiell^2}\nonumber\\ &&
-\frac{12\etaell(4\etaell^2\xiell^4+3\xiell^6-7\etaell^2\xiell^2-7\xiell^4+3\etaell^2+4\xiell^2)F_{1, \xiell}}{(\etaell^2-\xiell^2)^2(1-\etaell^2)^2\xiell}\nonumber\\&&
+\frac{12\etaell (2\etaell^6\xiell^2-4\etaell^4\xiell^4+5\etaell^2\xiell^6+3\xiell^8-5\etaell^4\xiell^2-2\etaell^2\xiell^4-5\xiell^6+3\etaell^4+3\etaell^2\xiell^2)F_1}{ (\etaell^2-\xiell^2)^3(1-\etaell^2)^2\xiell^2},\nonumber \eea
\bea {C}_2 &=&\frac{12\xiell(\xiell^2-1) F_{2,\xiell \xiell}}{(1-\etaell^2)(\etaell^2-\xiell^2)}+\frac{12\etaell F_{2, \xiell \etaell}}{\etaell^2-\xiell^2}
-\frac{4(\etaell^4+7\etaell^2\xiell^2+4\xiell^4-6\etaell^2-6\xiell^2)F_{2,\xiell}}{(\etaell^2-\xiell^2)^2(1-\etaell^2)}\nonumber\\ &&-\frac{12\xiell\etaell(\etaell^2+\xiell^2-2)F_{2, \etaell}}{(\etaell^2-\xiell^2)^2(\xiell^2-1)} +\frac{4\xiell\left(\etaell^6-8\etaell^4\xiell^2+\etaell^2\xiell^4+11\etaell^4-4\etaell^2\xiell^2
+5 \xiell^4-6\etaell^2 \right)F_2}{(\etaell^2-\xiell^2)^3(1-\etaell^2)(\xiell^2-1)}\nonumber\\ &&
+\frac{12(\etaell^4+2\etaell^2\xiell^2-2\etaell^2-\xiell^2)F_{4,\xiell\etaell}}{(\etaell^2-\xiell^2)(\xiell^2-1)\etaell}\nonumber\\ &&
-\frac{12(\etaell^8+4\etaell^6\xiell^2+2\etaell^4\xiell^{4}-4\etaell^6-8\etaell^4\xiell^2-2\etaell^2\xiell^4+4\etaell^4+2\etaell^2\xiell^2+\xiell^4)F_{4, \xiell}}{(\etaell^2-\xiell^2)^2(1-\etaell^2)(\xiell^2-1)\etaell^2}\nonumber\\ &&
-\frac{12\xiell(3\etaell^6+4\etaell^{4}\xiell^2-7\etaell^4-7\etaell^2\xiell^2+4\etaell^2+3\xiell^2)F_{4, \etaell}}{(\etaell^2-\xiell^2)^2(\xiell^2-1)^2\etaell}\nonumber\\&&
-\frac{12\xiell (3\etaell^8+5\etaell^6\xiell^2-4\etaell^4\xiell^4+2\etaell^2\xiell^6-5\etaell^6-2\etaell^4\xiell^2-5\etaell^2\xiell^4+3\etaell^{2}\xiell^2+3\xiell^4)F_4}{(\etaell^2-\xiell^2)^3(\xiell^2-1)^2\etaell^2}.\nonumber \eea
Here the functions $F_j$ are defined as $F_j=\hat{F}_j/{(\etaell^2-\xiell^2)^3}$, where
\bea
\hat{F}_1&=&{(1-\etaell^2)^{3/2}(\xiell^2-1)^{3/2}\left[
\xiell^3A_{300}+\etaell \xiell^2 A_{201}+\etaell^2\xiell A_{102}+\etaell^3 A_{003}\right]}\nonumber\\ &&
+{(1-\etaell^2)^{5/2}(\xiell^2-1)^{1/2}\left[\xiell^3 A_{120}+\etaell \xiell^2 A_{021}\right]}\nonumber\\
&&-{(1-\etaell^2) ^2(\xiell^2-1)\left[
\xiell^3 A_{210} +\etaell \xiell^2 A_{111}+\etaell^2\xiell A_{012}\right]}-{(1-\etaell)^3\xiell^3A_{030}}\nonumber\\
\hat{F}_2&=&-{(1-\etaell^2)^{3/2}(\xiell^2-1)^{3/2}\left[
3\etaell\xiell^2A_{300}+\xiell(2\etaell^{2}+\xiell^2)A_{201}+\etaell(\etaell^2+2\xiell^2)A_{102}+3\etaell^2\xiell A_{003}\right]}\nonumber\\ &&
+{(1-\etaell^2)^{3/2}(\xiell^2-1)^{1/2}\left[(\etaell^2+2\xiell^2-3)\etaell\xiell^2 A_{120}+\xiell (3\etaell^2\xiell^2-2\etaell^2-\xiell^2) A_{021}\right]}\nonumber\\
&&-(1-\etaell^2)(\xiell^2-1)\left[
\xiell^2\etaell(2\etaell^2+\xiell^2-3) A_{210} +\xiell(\etaell^4+2\etaell^2\xiell^2-2\etaell^2-\xiell^2)A_{111} \right.\nonumber\\ && \left. +\etaell (3\etaell^2\xiell^2-\etaell^2-2\xiell^2)A_{012}  \right]-3{(1-\etaell)^2(\xiell^2-1)\etaell\xiell^2A_{030}}\nonumber\\
\hat{F}_3&=&{(1-\etaell^2)^{3/2}(\xiell^2-1)^{3/2}\left[
3\etaell^2\xiell A_{300}+\etaell(\etaell^2+2\xiell^2)A_{201}+\xiell(2\etaell^2+\xiell^2)A_{102}+3\etaell\xiell^2 A_{003}\right]}\nonumber\\ &&
+{(1-\etaell^2)^{1/2}(\xiell^2-1)^{3/2}\left[(2\etaell^2+\xiell^2-3)\etaell^2\xiell A_{120}+\etaell (3\etaell^2\xiell^2-\etaell^2-2\xiell^2) A_{021}\right]}\nonumber\\
&&+(1-\etaell^2)(\xiell^2-1)\left[ \etaell^2\xiell(\etaell^2+2\xiell^2-3) A_{210} +\etaell(\xiell^4+2\etaell^2\xiell^2-\etaell^2-2\xiell^2)A_{111} \right. \nonumber\\&& \left.+\xiell(3\etaell^2\xiell^2-2\etaell^2-\xiell^2)A_{012}\right]
-3{(1-\etaell^2)(\xiell^2-1)^2\etaell^2\xiell A_{030}}\nonumber\\
\hat{F}_4&=&-{(1-\etaell^2)^{3/2}(\xiell^2-1)^{3/2}\left[
\etaell^3A_{300}+\etaell^2 \xiell A_{201}+\etaell\xiell^2 A_{102}+\xiell^3 A_{003}\right]}\nonumber\\ &&
-{(1-\etaell^2)^{1/2}(\xiell^2-1)^{5/2}\left[\etaell^3 A_{120}+\etaell^2 \xiell A_{021}\right]}\nonumber\\
&&-{(1-\etaell^2) (\xiell^2-1)^2\left[
\etaell^3 A_{210} +\etaell^2 \xiell A_{111}+\etaell\xiell^2 A_{012}\right]}-{(\xiell^2-1)^3\etaell^3A_{030}}\nonumber
\eea
As in the case of the separation in the parabolic coordinates, we can see immediately from the form of the linear compatibility condition \eqref{linearcompell} that both of the functions $W_1$ and $W_2$ will have only non-movable singularities unless either $F_2\equiv 0$ or $F_3\equiv 0.$ However, since both of the functions depend on all of the $A_{jk\ell}$'s and will vanish only if all of the coefficients do, there will be no such solution. In other words, if any of the $A_{jk\ell}$'s are not 0 then both $W_1$ and $W_2$ will satisfy linear determining equations and so will have only non-movable singularities.

Let us conclude this section with a discussion of the solutions of the linear determining equations \eqref{linearcompell}. In each of the previous cases, the coupled ODEs were separated by taking a sufficient number of derivatives to remove one of the functions from the equation.  However, the linear determining equations in the elliptic case have coefficients that are not rational  functions and so this method breaks down.
One may try to get rid of the square roots by isolating them on the right-hand side and then squaring the equation.
However, although we get rid of the square roots in this way, we have to deal with a nonlinear equation, that not surprisingly turns out to be even more complicated than the original condition \eqref{linearcompell}. A fruitful direction seems to be to follow the method that was initially used by Gravel\cite{Gravel}, namely to evaluate the linear compatibility condition at a  supposedly regular point, for example $x=0$ and $y=0$ in Cartesian coordinates and $\etaell,\xiell=\pm 1$ in elliptic. Of course, these points may well happen to be singular points of the potential, as was the case in Cartesian coordinates, possibly leading to  some solutions being lost. However, in the elliptic case even this simplification leads to systems of ODEs that are difficult to solve. Still the classification of third-order superintegrable systems that admit separation of variables in elliptic coordinates remains an open problem.

\section{Conclusion and Open Problems}
In this article, we consider the problem of a Hamiltonian system that admits separation of variables in orthogonal coordinates
in two-dimensional Euclidean space as well as another integral of degree three in the momentum. For systems that separate in
subgroup-type coordinates, several new families of such potentials were  already identified\cite{Gravel, GW, TW20101}.
There are potentials that satisfy linear determining equations, for example the 3:1 oscillator, as well as potentials that depend on nonlinear special functions, such as the Painlev\'e transcendents. In section 2, we have reviewed the conditions necessary for the existence of such nonlinear potentials. Namely, the requirement that one of the determining equations for the potential, the so-called linear compatibility condition either  vanishes or depends only on a function of one variable.
For subgroup-type coordinates, this condition can be satisfied for certain non-zero values of the coefficients $A_{jk\ell}$.
On the other hand, for systems that separate in parabolic coordinates\cite{PopperPostWint2012} the linear compatibility
condition will be independent of $W_1(\xi)$ or $W_2(\eta)$ if and only if all the $A_{jk\ell}$ are 0, leaving a first-order integral instead of a third. The main result of this paper is to finish the classification with a similar statement for the elliptic coordinate case, where it has been shown that all potentials will also have only non-movable singularities and so will not depend on nonlinear special functions.

This leads to at least two interesting open problems and conjectures. The first open problem is to solve the linear determining equations for the elliptic case and to classify all systems with a third-order integral that admit separation of variables in elliptic coordinates. In the parabolic case, all of the obtained potentials admitted an additional second-order integral of motion, so that the third-order integral was the Poisson or Lie bracket of the two second-order ones, neither being the Hamiltonian. We conjecture that this is also the case in elliptic coordinates so in particular there are no new  superintegrable Hamiltonians separable in the elliptic coordinates possessing a third order integral of motion.

The second open problem is to explain the difference between the two pairs of cases. Why should it be that subgroup type coordinates are necessary to  allow superintegrable potentials that depend on nonlinear special functions? Furthermore, the lack of new solutions in the parabolic and, if true, in the elliptic case is striking and the connection with the type of orthogonal separation is not clear. It is  interesting to note that in 2D both subgroup coordinates have at least one ignorable coordinate (i.e. a coordinate that does not appear in the metric)  whereas the non-subgroup type coordinates don't. In higher dimensions, this is  no longer the case and in particular there exist coordinate systems with an ignorable coordinate that are not of subgroup type\cite{miller1981subgroups}. Thus extending the classification to higher dimensions may help to clarify the mechanism for existence of such ``nonlinear" potentials.

\begin{acknowledgments}
The research of A. M. was supported by the European social fund within the framework of realizing the project ``Support of inter-sectoral mobility and quality enhancement of research teams at Czech Technical University in Prague'', CZ.1.07/2.3.00/30.0034.

L. \v{S}. was supported by the Grant Agency of the Czech Technical University in Prague, grant No. SGS 13/217/OHK4/3T/14 and by the Czech Ministry of Education, RVO68407700. 

S. P. acknowledges funding from the Vice Chancellor of Academic Affairs and the College of Natural Science of U. Hawai`i at M\=anoa for support.

The authors thank Pavel Winternitz for intense discussions on the subject of this paper.

\end{acknowledgments}

\bibliography{bib}
\bibliographystyle{unsrt}
\end{document}